# EVALUATION OF NEUROCOMBAT AND DEEP LEARNING HARMONIZATION FOR MULTI-SITE MAGNETIC RESONANCE NEUROIMAGING IN YOUTH WITH PRENATAL ALCOHOL EXPOSURE

*Chloe Scholten[1*], Elyssa M. McMaster[2*], Adam M. Saunders[2], Michael E. Kim[2], Gaurav Rudravaram[2], Elias Levy[2], Bryce Geeraert[1], Lianrui Zuo[2], Simon Vandekar[3], Catherine Lebel[1], Bennett A. Landman[2,4]*

[1]Cumming School of Medicine, University of Calgary, Calgary, AB, Canada, [2]Vanderbilt University, Nashville, TN, USA, [3]Vanderbilt University Medical Center, Nashville, TN, USA * *Indicates equal contribution*

## ABSTRACT

In cases of prevalent diseases and disorders, such as Prenatal Alcohol Exposure (PAE), multi-site data collection allows for increased study samples. However, multi-site studies introduce additional variability through heterogeneous collection materials, such as scanner and acquisition protocols, which confound with biologically relevant signals. Neuroscientists often utilize statistical methods on image-derived metrics, such as volume of regions of interest, after all image processing to minimize site-related variance. HACA3, a deep learning harmonization method, offers an opportunity to harmonize image signals prior to metric quantification; however, HACA3 has not yet been validated in a pediatric cohort. In this work, we investigate HACA3's ability to remove site-related variance and preserve biologically relevant signal compared to a statistical method, neuroCombat, and pair HACA3 processing with neuroCombat to evaluate the efficacy of multiple harmonization methods in a pediatric (age 7 to 21) population across three unique scanners with controls and cases of PAE with downstream MaCRUISE volume metrics. We find that HACA3 qualitatively improves inter-site contrast variations, but statistical methods reduce greater site-related variance within the MaCRUISE volume metrics following an ANCOVA test, and HACA3 relies on follow-up statistical methods to approach maximal biological preservation in this context.

## 1. INTRODUCTION

Prenatal Alcohol Exposure (PAE) occurs in 1 in 10 pregnancies worldwide and results in structural brain changes that yield cognitive and behavioral challenges [1], [2]. Structural magnetic resonance imaging (MRI) provides a non-invasive method to image pathology-driven neurological alterations in vivo [3]. The manifestations of PAE vary with the dose and timing of alcohol exposure, postnatal environment, and genetic features, which induce variability in PAE cases [1], [2] and establish the need for large sample sizes to make population-level conclusions about PAE.

Multi-site MRI acquisition offers a solution for large-scale data collection, but introduces heterogeneity in image contrast, especially in cases without a common protocol or scanner model (Fig. 1). neuroCombat, a statistical harmonization approach to estimate site-specific biases and scaling effects, is the current standard among neuroscientists for metrics derived after image processing [4]. Although neuroCombat exhibits limitations in settings with small per-iste sample sizes or non-normal distributions [5], it has been shown to effectively capture population-level site biases in neuroimaging data [4]. In such settings, complementary harmonization strategies may be necessary.

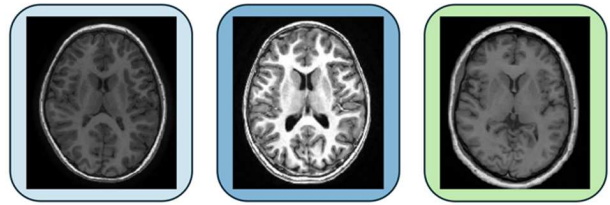

| Site | Calgary Pre-Upgrade | Calgary Post-Upgrade | Edmonton |
|---|---|---|---|
| Scanner | GE 3T MR750w | GE 3T MR750w UHP | Siemens 3T Prisma |
| Sequence | BRAVO | BRAVO | MP-RAGE |
| Resolution | 0.8 mm³ | 0.8 mm³ | 0.9 mm³ |
| TR (ms) | 8.2 | 8.2 | 1800 |
| TE (ms) | 3.2 | 3.5 | 2.37 |
| Flip Angle | 10° | 8° | 8° |
| TI (ms) | 600 | 600 | 900 |

**Fig. 1.** Heterogeneous site protocols invite site-related variation that confounds underlying biological variation.. Calgary Pre-Upgrade was a GE 3T MR750w scanner, Calgary Post-Upgrade was a GE 3T MR750w UHP scanner, and Edmonton was a Siemens 3T Prisma. Resolutions were isotropic at all sites.

Deep learning, image-based methods for harmonization show promise for heterogenous, pathological datasets like PAE. Harmonization with Attention-based Contrast, Anatomy, and Artifact Awareness (HACA3) is an approach that utilizes deep learning to harmonize data with various MR contrasts from multiple sites to a target contrast while preserving anatomical detail [6]. As input, HACA3 accepts T1-weighted (T1w), T2-weighted (T2w), Proton Density-weighted (PDw), and FLAIR images from a subject and another subject's image with the target contrast. The HACA3 algorithm leverages anatomical information to represent subject images as though they had been acquired with a different protocol. After being tested on both healthy subjects as well as people with multiple sclerosis, HACA3 showed broad applicability in control and pathological adult datasets [6]. Despite that HACA3 outperforms statistical harmonization method in harmonizing adult brain MRI data [7], it has yet to be tested in a pediatric population. In this work, we investigate HACA3's potential to remove site-related effects within T1w images in a PAE, pediatric-specific study. Additionally, this study provides an evaluation in a pediatric PAE cohort comparing an image-level deep learning harmonization method (HACA3) and a metric-level statistical harmonization (neuroCombat), with practical guidance on when each is appropriate. We hypothesize that image contrast harmonization with HACA3 paired with neuroCombat reduces site biases and allows for greater differentiation between PAE and non-PAE groups when compared to pre-harmonization data or harmonization with neuroCombat or HACA3 alone.

## 2. METHODOLOGY

The study cohort includes 371 MRI images (177 T1w, 168 T2w, 26 PDw) from 131 adolescents (63 with prenatal alcohol exposure; 59 females) aged 7-21 years. The data came from three sites with differing scanners and protocols: Calgary Pre-Upgrade, Calgary Post-Upgrade, and Edmonton. All sites collected T1w and T2w images; only Edmonton collected PDw images. T2w and PDw images were only used as input for HACA3 to provide all available anatomical information; only T1w images are used in downstream analysis.

All images underwent preprocessing to prepare for HACA3 harmonization (Fig. 2). T1w images were preprocessed with N4 Bias Correction [8]. T1w images for neuroCombat evaluation were resampled to 1mm$^3$ for common voxel spacing despite different protocols [9]. Non-T1w HACA3 inputs underwent super-resolution correction 2D acquisitions [10]. All HACA3 input images were registered to an MNI [11] atlas with 0.8mm$^3$ resolution. These preprocessing steps were consistent with the original HACA3 validation paper to ensure consistency [6]. The HACA3 model's weights for this experiment are available at https://github.com/lianruizuo/haca3. Notably, the published model weights have only been trained on adult data [6].

HACA3 inputs included all T1w, T2w, and PDw images acquired for each subject. The outputs from HACA3 were 177 images harmonized to one subject's T1w contrast from the Calgary post-upgrade cohort, including the remaining Calgary post-upgrade subjects. Output T1w data were processed through SLANT-TICV [12] and MaCRUISE [13], [14] in a reproducible manner [15] for regional segmentation both with and without HACA3 harmonization. Brain volumes in 5 regions bilaterally (anterior cingulate gyrus, amygdala, hippocampus, medial frontal cortex, and superior frontal gyrus) and total intracranial volume were extracted. These regions were chosen as they have previously shown reductions in adolescents with PAE [16]. Thirteen scans were excluded at this stage due to poor MaCRUISE segmentations upon visual inspection.

All statistical analyses were completed in RStudio (v4.2.1, R Core Team 2021). A total of 164 scans were included: Calgary Pre-Upgrade (n=87 with 47 PAE); Calgary Post-Upgrade (n=45 with 14 PAE), Edmonton (n=32 with 12 PAE). ANCOVA tests were used to quantify site effects in each harmonization method (non-harmonized, neuroCombat, HACA3, HACA3 + neuroCombat) for each brain region after controlling for sex, age, PAE status, and total intracranial volume; ComBat has been validated for longitudinal studies [17]. Type III sum-of-squares ANOVAs were performed to extract F-statistics and p-values. To determine preservation of biological features after harmonization, linear mixed effects models were used to test the effect of PAE on regional volumes while controlling for age, sex, total intracranial volume and including a random intercept for participant. FDR correction was used to correct all p-values for multiple comparisons separately for each harmonization type. The effect size and level of significance were compared between the non-Harmonized data, and data harmonized with neuroCombat, HACA3, or HACA3 + neuroCombat.

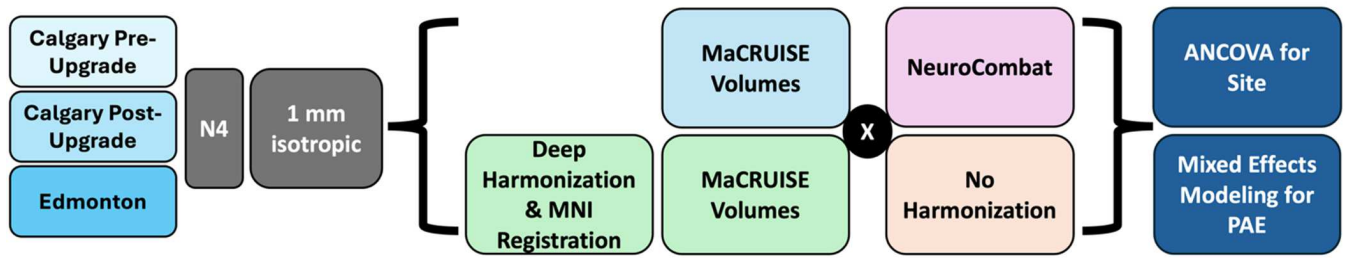

**Fig. 2.** Images underwent preprocessing and deep learning harmonization (HACA3) in two different pathways followed by common statistical tests to test our hypothesis that a combination of image-based and statistical-based harmonization methods yield optimal mitigation of site effects.

## 3. RESULTS

Contrast differences between sites visually appeared reduced following deep learning harmonization with HACA3 in a qualitative evaluation (Fig. 3). In the non-harmonized method, statistically significant effects of site were observed in the left anterior cingulate gyrus, the right anterior cingulate gyrus, the left amygdala, the left medial frontal cortex, the right medial frontal cortex, the left superior frontal gyrus and the right superior frontal gyrus. neuroCombat harmonization eliminated all significant site effects. In data only harmonized with HACA3, there were significant site effects in the same regions as non-harmonized data. Performing neuroCombat on the data harmonized with HACA3 eliminated all previously mentioned significant site effects (Fig. 4).

The effects of PAE on regional brain volumes were tested in the four harmonization methods. Regardless of the harmonization method, we found significantly reduced total intracranial volume in the PAE group (avg. $\beta = -6.28 \times 10^4$, avg. $p<0.01$). In non-harmonized and neuroCombat harmonized models, the PAE group had reduced volumes in the right superior frontal gyrus (Non-Harmonized: $\beta = -8.37 \times 10^2$, p=0.034; neuroCombat: $\beta = -7.92 \times 10^2$, p=0.035), but these relationships were not seen in data that been harmonized using HACA3 ($\beta = -7.72 \times 10^2$, p=0.17) or HACA3 + neuroCombat ($\beta = -7.17 \times 10^2$, p=0.20) (Fig. 5).

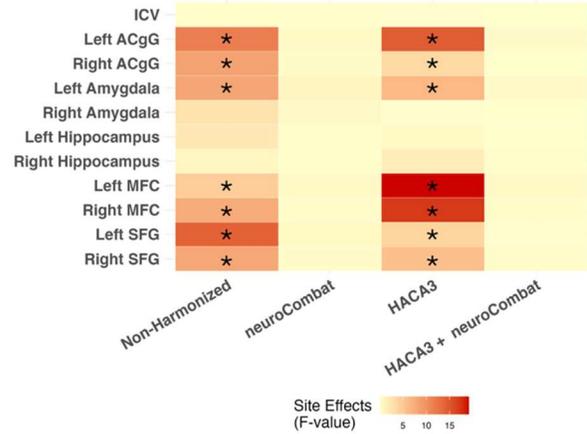

**Fig. 4.** ANCOVA tests were used to compare mean regional volume at each site in each harmonization method while adjusting for age, sex, PAE status, and ICV. F-statistics and p-values were extracted from Type III ANOVA tables, and p-values were corrected for multiple comparisons using FDR. Larger F-values (and smaller p-values) indicate a larger difference between mean thus stronger site effects. Site effects are reduced in volumes that were harmonized with neuroCombat. Asterisks indicate significant effect ($p<0.05$).

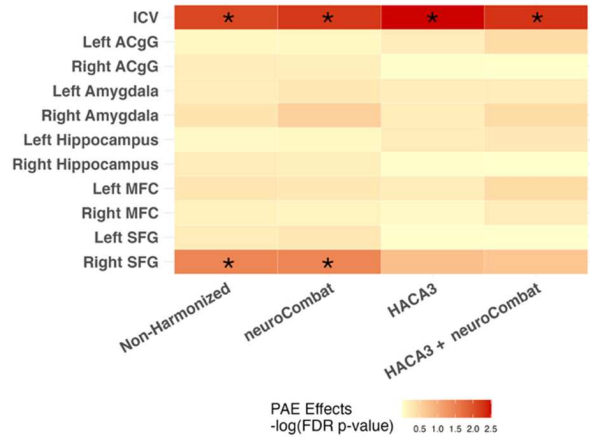

**Fig 5.** Linear mixed effects models were used to test the effect of PAE on regional brain regions in the different harmonization methods while controlling for age, sex, PAE status and ICV. P-values were corrected for multiple comparisons using FDR. Darker colours indicate a stronger effect (smaller p-value). Asterisks indicate significant effect ($p<0.05$).

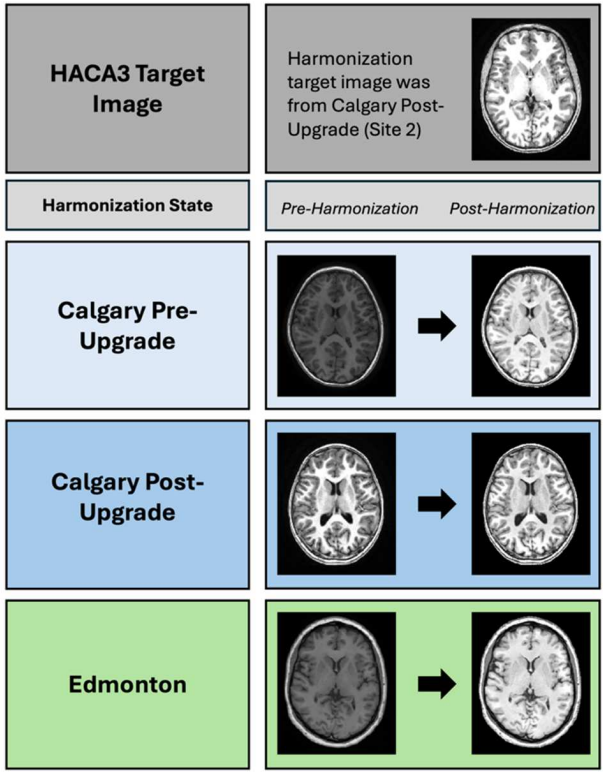

**Fig. 3.** Despite a common intensity range for all images, we observe visual qualitative contrast differences between pre- and post-HACA3 images.

## 4. CONCLUSION

This work provides a systematic evaluation of image-level and metric-level harmonization methods in a multi-site pediatric neuroimaging cohort. We compare a deep learning–based approach (HACA3) and a statistical method (neuroCombat) with respect to site-effect removal and preservation of biological group differences and examine how harmonization at different stages of the pipeline affects downstream volumetric analyses. We show that HACA3 harmonization qualitatively reduced contrast variations due to scanner, though quantitative brain volume metric agreement improves when paired with a statistical method. This may highlight the need to fine-tune HACA3 (which is currently trained on adults only data) on pediatric images. HACA3 has great potential to harmonize image contrast in pediatrics and further research is needed in pediatric-specific image harmonization.

## 5. ACKNOWLEDGEMENTS

This work was supported by funding from the Canadian Institute of Health Research (CIHR), the Natural Sciences and Engineering Research Council of Canada (NSERC), and the Addictions & Mental Health Strategic Clinical Network. CS receives scholarship support from the Canadian Institute of Health Research (CIHR), the Canadian Neurodevelopment Research Training Platform (Can-NRT), and the Hotchkiss Brain Institute (HBI). CL receives salary support from the Canada Research Chairs Program.

This work has been funded in part by NIH 5U01DA055347-03, NIH 1R01EB017230, U24AG074855, P50HD103537, U01CA152662 (Grogan), R01CA253923 (Landman & Maldonado), R01 CA275015 (Maldonado & Lenburg), and U01 CA196405 (Maldonado). This work was supported by the Alzheimer's Disease Sequencing Project Harmonization Consortium (ADSP-PHC) that is funded by NIA (U24 AG074855, U01 AG0668057 and R01 AG059716). This work was supported by the Alzheimer's Disease Sequencing Project Phenotype Harmonization Consortium (ADSP-PHC) that is funded by NIA (U24 AG074855, U01 AG068057 and R01 AG059716). This work was also supported by U01CA152662, R01CA253923, R01CA275015, and U01 CA196405. The content is solely the responsibility of the authors and does not necessarily represent the official views of the NIH.

## 6. COMPLIANCE WITH ETHICAL STANDARDS

Data collection was approved in Calgary at the Alberta Children's Hospital (REB17-0662) and in Edmonton at Peter S Allen MRI Centre (Pro00093230).